\documentclass[10pt,a4paper]{article}
\usepackage{amsfonts,graphicx}
\usepackage{amsmath, amsthm}
\usepackage{latexsym}
\usepackage{subfigure} %for joining figures together
\usepackage{verbatim}
\usepackage{amssymb}
\usepackage{mathrsfs}
\usepackage{hyperref}
\usepackage{bm}

\newtheorem{proposition}{Proposition}[section]
\newtheorem{lemma}{Lemma}[section]
\newtheorem{corollary}{Corollary}[section]
\newtheorem{example}{Example}[section]

\title{Robust Dynamic Selection of Tested Modules in Software Testing Process for Maximizing Delivered Reliability}
\author{ Ping Cao$^{a,}$\footnote{Corresponding author. Tel. number: +86 13167731296. Email address: pcao@ustc.edu.cn.}, Zhao Dong$^b$,
Ke Liu$^b$,
Kai-Yuan Cai$^{c,d}$
\\$^a$\footnotesize{School of Management, University of Science and Technology of China, Hefei, 230026, China}
\\$^b$\footnotesize{MADIS and National Center for Mathematics and Interdisciplinary Sciences,} \\ {\footnotesize Academy of Mathematics and Systems
Sciences, CAS, Beijing  100190, China}\\
$^c$\footnotesize{Department of Automatic Control, Beijing University of Aeronautics}\\
\footnotesize{and Astronautics, Beijing 100191, China} \\
$^d$\footnotesize{State Key Laboratory of Computer Science, Institute of Software, CAS, Beijing 100190, China}\\
}
\date{}

\begin{document}
\maketitle

\begin{abstract}
Software testing is aimed to improve the delivered reliability of the users. Delivered reliability is the reliability when using the software after it is delivered. Since the software consists of many modules, the delivered reliability depends on the number of defects remaining in each module as well as the users' operational profile, which specifies how the users will use these modules. A good testing policy should take the users' operational profile into account and dynamically select tested modules according to the current state of the software during the testing process. This paper discusses how to dynamically select tested modules in order to maximize delivered reliability by formulating the selection problem as a dynamic program. Since the testing process is performed only once, the tester is more concerned about the realized reliability rather than its expected value. Hence, the realized risk of reliability must be considered during the testing process, which is described by the tester's utility function in this paper. Besides, since usually the tester has no accurate estimate of the users' operational profile, by employing robust optimization technique, we analyze the selection problem in the worst case given that the operational profile lies in a given uncertainty set. By numerical examples, we show the necessity of taking maximizing delivered reliability directly as the objective and using robust optimization technique when the tester does not know the detailed value of operational profile. Moreover, it is shown that the risk averse behavior of the tester has a great impact on the distribution of delivered reliability.
\end{abstract}

\section{Introduction}

Software testing is the most popular method for improving the software reliability~\cite{p01}. It is a major paradigm for software quality
assurance and is extensively carried out in nearly every
software development project~\cite{p00}. During the software testing process, the defects remaining in the software will be detected and removed subsequently, resulting in the improvement of software reliability.

A typical software testing process can be depicted in the following way~\cite{c07}: At the beginning the test cases are generated and subsequently divided into several classes. For each testing, first a class of test cases is selected, and then a test case will be chosen randomly from this class. Executing this test case will result in either detecting a defect or detecting no defect. In the former case, the detected defect will be removed subsequently. Then a test is completed and the tester moves to the next test. In the latter case, the tester moves to the next test directly. We can control the testing process by appropriately selecting the classes. Several software testing methods are proposed based on special selection patterns, including Markov usage model based testing and random testing~\cite{wt94}. Markov usage model based testing assumes that the classes will be selected according to a Markov chain and random testing assumes that the classes will be selected randomly. However, these methods all assume that the selection process has no relationship with the testing result. Thus, these methods may not be optimal to detect and remove the defects remaining in the software, which may lower the effectiveness of the testing strategy. Therefore, a good testing strategy should take the current state of the software into account.

There are two main goals in software testing~\cite{fhls1998}. One goal is from the tester's perspective, detecting the defects so that they can be removed and thus the reliability is improved. The other goal is from the users' perspective, evaluating users' delivered reliability to gain confidence that the software is sufficiently reliable for its use. Here, delivered reliability is the reliability of using the software after it is delivered. A tester may try to detect all latent defects and remove these defects as many as possible. However, due to limited resources, including time and labor resources, it is often impossible to remove all defects. Note that different defects contribute differently to the software's delivered reliability for different users, which can be characterized by the users' operational profile. Thus, a natural problem arises: How to choose testing strategy to improve the reliability from the users' perspective, or the delivered reliability?

A lot of literature in software testing studies how to remove the defects as many as possible during the testing process~\cite{c01,c02,c03,c05}, and the effect of operational profile on the delivered reliability is often ignored. However, the actual goal of software testing is not only removing the defects but also improving delivered reliability. Thus, the ignorance will make the proposed testing strategy less effective than one expected. It is quite necessary to directly set maximizing delivered reliability as the objective. Besides, although there is a qualitative relationship between the remaining defects and delivered reliability, the quantitative relationship is hard to depict. Many papers including~\cite{c07,c08,c10} discuss how to assess or evaluate the delivered reliability given a testing policy. However, they assume that the testing policy is a given Markov usage model based testing and ignore the fact that the testing policy may vary with the testing process. As far as the authors know, there is no paper studying how to achieve a high delivered reliability by dynamically selecting the test cases during the testing process. But it is a topic worth discussing. Removing the defects is not the ultimate goal of software testing. We test the software to make it work well and meet end-user's needs. Therefore, improving delivered reliability is much more closed to the tester's purpose than simply removing defects. It is quite necessary to make the tester's objective be maximizing the delivered reliability rather than minimizing the number of remaining defects. In fact, we can show that minimizing the number of remaining defects may result in a significant small delivered reliability.

One obstacle that hinders the tester from taking maximizing the delivered reliability as his testing goal is that the delivered reliability is related to the users' operational profile which is often unknown during the software testing process. Operational profile describes how the users will use the software, i.e., in which frequency the users will use each module of the software.  A lot of literature assumes that the operational profile is known before testing\cite{os03,yx00}. This assumption seems unreasonable, but it can be explained by that the tester may have the operational profiles of several softwares with similar function. These profiles can be used to estimate the operational profile of the software under test. It is highly probable that the estimated operational profile is quite different from the real one. A wrongly assumed operational profile will lead to a selection policy quite deviated from the optimal one, which will result in a lower delivered reliability. In this paper, since we take maximizing the delivered reliability as our testing goal, the uncertainty of operational profile must be taken into account. By assuming that the operational profile lies in an uncertainty set, we employ robust optimization technique to find a selection policy that works well even in the worst case. Our numerical example shows that the resulting selection policy based on robust optimization has a stable performance, no matter what the true value of the operational profile is.

A significant difference between the software testing process and other stochastic decision processes is that it is performed only once, i.e., the process is unrepeatable. Thus, the objective which is to maximize expected reliability at the end of the testing process may not be adequate in this situation as a single poor realization can have a great impact on the reliability. Thus, the tester may want to control the risk during the testing process and he may be willing to use selection polices that sacrifice some expected reliability in return for a less risk, or volatility of the reliability. In other words, the tester is prone to be risk averse rather than risk neutral. In this paper, we capture the tester's risk averse behavior by using a strictly increasing and concave utility function.

There is a lot of literature discussing how to dynamically select the test classes. Cai has done a lot of work on the controlled Markov chain approach to software testing~\cite{c01,c02,c03,c05}. Those papers study how to test software such that all the remaining defects are detected and removed at the least expected cost or how to test software such that defects are detected and removed as many as possible given the expense budget or the maximal number of test times. Therefore, the objective in those papers is to minimize the number of the remaining defects rather than to maximize the delivered reliability. Moreover, those papers do not incorporate the risk into the testing process. However, these issues cannot be ignored in the practical testing process.

This paper studies how to dynamically select the classes during the software testing process to maximize delivered reliability. Our contribution is three fold. First, the objective becomes maximizing delivered reliability. Second, we consider the worst case when the tester faces uncertainty of the operational profile. Usually the tester cannot obtain the exact value of the users' operational profile, since different software may have different operational profile and the tester may not be able to predict accurately the operational profile before the software is released to the market. However, the tester may have a rough estimate by statistical inference from the operational profiles of softwares with similar function. Therefore, we can assume that the value of the users' operational profile lays in a given uncertainty set. Under this assumption we can discuss this problem by robust optimization method. Finally, we consider the risk factor in the selection process. In most of the literature the risk factor is ignored and the average performance of the selection policy is the criterion. This may contradict with the common sense that the tester is prone to be risk averse as for a specified software the testing process will be performed only once. Therefore, the expected performance may be quite different with the realized performance. In this sense, the risk must be taken into account.

It is worth mentioning that there are also several papers discussing the risk during the testing, such as~\cite{a00,b99}. However, the risk mentioned in this paper is quite different from that in~\cite{a00,b99}. In~\cite{a00,b99}, risk means something that might happen. Thus, it can be measured by the probability of a fault. The risk in those papers is quite similar with the reliability except that it also takes the cost of a fault into account. In this paper, risk is the deviation of the realized reliability from the expected reliability, which can be called measure risk. Therefore, they are two different concepts of risk.

This paper is structured as follows. In Section 2, we describe our software testing model and formulate it as a dynamic program. In Section 3, we analyze the optimal utility function and the optimal selection policy. In Section 4, we conduct several numerical examples to examine the impact of testing objective, the necessity of considering the uncertainty of operational profile, and the effects of tester's risk aversion behavior. Section 5 concludes this paper.

\section{Model Description and Mathematical Formulation\label{se_2}}

%We take the following assumptions for the software testing process and the software defect removal mechanism.

Consider a finite horizon software testing process to test a software with $m$ modules. At the beginning of the process, we divide the test cases into $m$ classes according to the $m$ modules. Each class $i$ is specialized in testing module $i$ and $\mathcal{C}_i$ is the set of all possible test cases to the corresponding class $i$. For a given module $i$, let $N_i$ be the number of defects\footnote{In practice, the initial number of defects in module $i$ is unknown. However, since the topic of this paper is focused on evaluating the impact of operational profile on the selection process, we assume that $N_i$ is known in this paper.}, $i=1,2,\cdots, m$. After the execution of a test case, a defect is found only if it triggers a failure which is observed from the test result. The test result may reveal many failures, but one failure only corresponds to one defect (see~\cite{kpay11} for similar assumptions). We assume that the occurrence of the possible failures are independent, and hence each defect in module $i$ can be found independently\footnote{As mentioned in~\cite{d88}, the time to detect a defect seems to be nearly exponentially distributed and the scale parameter is stable across modules. Therefore, it is reasonable to assume that the defects in module $i$ are independent with each other and each defect in $\mathcal{C}_i$ will trigger a failure with a probability independent of the time period.} with probability $\theta_i$\footnote{The parameter $\theta_i$ is in fact unknown. According to~\cite{d88}, a prior on $\theta_i$ can be used and updated according to the testing process. However, if we consider the updating  process for the parameter $\theta_i$,  the mathematical arguments will be quite messy. Since the subject of this paper is not focused on estimating these parameters,
we assume that $\theta_i$ is already known.} if a test case from class $i$ is executed, $i=1,2,\cdots,m$. Let $\mathbf{N} = (N_1,N_2,\cdots , N_m)$ and $\bm{\theta}=(\theta_1,\theta_2,\cdots,\theta_m)$. The software testing process lasts for $T$ periods\footnote{Usually there is a deadline such that the software should be released to the market before or at that time point. In this paper we assume that the testing times $T$ is exogenously given. However, this model can be readily extended to the case that the testing times can be dynamically adjusted.}, and the sequence of event in each period is as follows:
\begin{enumerate}
\item At the beginning of period $t$ ($t=0,1,\cdots, T-1$), the tester observes the system's state $\mathbf{N}(t)=(N_1(t), N_2(t), \cdots, N_m(t))$ and chooses module $i$ of the software to be tested.
\item The tester randomly select a test case from class $i$ and detects $W_i$ defects from the testing results. Since each defects in module $i$ will trigger a failure with probability $\theta_i$ independently and one failure corresponds to one defect, $W_i$ is a random variable following the binomial distribution with parameters $N_i(t)$ and $\theta_i$, i.e., $W_i\sim B(N_i(t),\theta_i)$.
\item The tester removes all the detected defects with probability 1 from module $i$\footnote{Here we assume that the defect removal is perfect, i.e., all the detected defects will be removed and no new defects are introduced. Imperfect defect removal can also be similarly considered with the model slightly modified.}.
\item The system state changes to
$$\mathbf{N}(t+1) = (N_1(t),N_2(t), \cdots, N_{i-1}(t), N_i(t)-W_i, N_{i+1}(t),\cdots, N_m(t)).$$
If $t+1=T$, then the software testing process ends and the software is released to the market. Otherwise, the system moves to period $t+1$.
\end{enumerate}

It is frequently assumed that after release the software will be used according to an operational profile which can be characterized by a probability vector $\mathbf{p}=(p_1,p_2,\cdots,p_m)'$, where the notation $'$ is used to transpose the matrix, $p_i\geq 0$, $i=1,2,\cdots,m$ and $\sum_{i=1}^mp_i=1$. Under the operational profile $\mathbf{p}$, the users will use module $i$ with probability $p_i$, $i=1,2,\cdots,m$ within a time period. Thus, if there are $x_i$ defects in module $i$ after release, $i=1,2,\cdots,m$ (letting $\mathbf{x}=(x_1,x_2,\cdots,x_m)'$), then the probability of using the software within a time period without triggering failures for the users is
\begin{equation}
R(\mathbf{x},\mathbf{p})= \sum_{i=1}^m p_i(1-\theta_i)^{x_i}, \label{eq_reliability}
\end{equation}
which can be used as a software reliability measure. Since the reliability is computed from the perspective of the users, $R(\mathbf{x},\mathbf{p})$ is called delivered reliability. In our model, the delivered reliability is $R(\mathbf{N}(T),\mathbf{p})$

There are a lot of software reliability measures in literature~\cite{t031}. In this paper we use the above reliability measure. However, the argument is similar for other reliability measures.

The tester's objective is to maximize the delivered reliability by dynamically selecting the tested modules at the beginning of each testing period. Note that there are two sources of uncertainty for the delivered reliability. One is the randomness of the system state $\mathbf{N}(T)$. The other is the uncertainty of the operational profile $\mathbf{p}$. For a new software under test which has not yet been released to the market and delivered to the users, it is hard, if not impossible, to estimate the exact value of the operational profile $\mathbf{p}$. However, by statistical inference from the historical records of softwares having similar functions with the software under test, it is reasonable to assume that $\mathbf{p}$ lies in an uncertainty set, denoted by $\mathcal{P}$, which is known to the tester.

In order to tackle the first source of uncertainty, a natural way is to maximize the expected delivered reliability instead. However, since the testing process is only performed once, the tester cares more about the realization of the reliability rather than its expectation. It may happen that the reliability induced by a selection policy is high in average while it can probably get a quite low value in its realization, which the tester wishes to avoid. Therefore, the tester is prone to be risk averse and he wants to control the reliability from below. Thus, we turn to maximize the expected utility of the delivered reliability, which is extensively adopted in the economic literature~\cite{al2008}. A typical risk-averse tester has a strictly increasing and concave utility function $U(\cdot)$. In order to tackle the second source of uncertainty, since we do not know the exact value of the operational profile, we turn to maximize the expected utility of the delivered reliability in the worst case given that the operational profile $\mathbf{p}$ lies in the known uncertainty set $\mathcal{P}$ instead, which is commonly adopted in the literature of robust optimization~\cite{bgn2009}.

It follows from the model description that  at the beginning of the software testing process, (i.e., at period $0$), $\mathbf{N}(0)=\mathbf{N}=(N_1,N_2,\cdots,N_m)'$. By the above argument, the dynamic selection problem can be formulated as
\begin{equation}\label{problem_def}
\max_{u\in\mathfrak{U}}\min_{\mathbf{p}\in \mathcal{P}}\mathbb{E}[U(R(\mathbf{N}(T),\mathbf{p}))|\mathbf{N}(0)=\mathbf{N}],
\end{equation}
where $\mathfrak{U}$ consists of all admissible selection policies, and $U(x)$ is a utility function which is increasing and concave in $x\in \mathbb{R}^+$.

Problem (\ref{problem_def}) can be solved by dynamic programming approach. We call the testing process is at state $(\mathbf{x},t)$ if the tester is  at the beginning of time period $t$ and there are $x_i$ defects remaining in module $i$, $i=1,2,\cdots,m$, where $\mathbf{x}=(x_1,x_2,\cdots,x_m)$. Define the optimal utility to go at state  $(\mathbf{x},t)$ by $$J_t(\mathbf{x})=\max_{u\in\mathfrak{U}}\min_{\mathbf{p}\in \mathcal{P}}\mathbb{E}[U(R(\mathbf{N}(T),\mathbf{p}))|\mathbf{N}(t)=\mathbf{x}].$$

Obviously, we have
\begin{equation}
J_T(\mathbf{x}) = \min_{\mathbf{p}\in \mathcal{P}} U(R(\mathbf{x},\mathbf{p})). \label{dp_boundary}
\end{equation}

Note that if the testing process is at state $(\mathbf{x},t)$ and the tester chooses to test module $i$, then the defect number in module $j$ will remain unchanged, for $j\neq i$. For each defect in module $i$, with probability $\theta_i$ it will be detected and removed. Thus, the number of defects remaining in module $i$ is a random number, which can be written as $\sum_{j=1}^{x_i}W_{i,j}$, where $W_{i,j}$ is independent with each other and follows the Bernoulli distribution with parameter $1-\theta_i$. Hence, the number of defects remaining in module $i$ follows the binomial distribution with parameters $x_i$ and $1-\theta_i$, $i=1,2,\cdots,m$.

By the principle of optimality, $J_t(\mathbf{x})$ satisfies the optimality equation
\begin{eqnarray}
J_t(\mathbf{x}) & = & \max_{1\leq i\leq m}\mathbb{E}J_{t+1}\left(x_1,x_2,\cdots,x_{i-1}, \sum_{j=1}^{x_i}W_{i,j}, x_{i+1},\cdots,x_m\right) \nonumber
\\ &=& \max_{1\leq i\leq m} \sum_{k=0}^{x_i}\left(\begin{array}{c}x_i\\k\end{array}\right) (1-\theta_i)^k \theta_i^{x_i-k} J_{t+1}(x_1,x_2,\cdots,x_{i-1}, k, x_{i+1},\cdots,x_m), \nonumber
\\ && \ \ \ \ \  \mbox{for}\ t=0,1,\cdots,T-1. \label{dp_recursion}
\end{eqnarray}

By the boundary condition (\ref{dp_boundary}) and the optimality equation (\ref{dp_recursion}), we can solve $J_t(\mathbf{x})$ and the optimal selection policy recursively in a backward manner, from $t=T$ to $t=0$.

\section{Analysis}

In this section, we give properties of the optimal utility function $J_t(\mathbf{x})$ and the optimal selection policy. These properties can help characterize the
optimal function and make the tester more clear about the relationship of the profit function $J_t(\mathbf{x})$ with the remaining defect numbers $\mathbf{x}$ and current testing period $t$.

\subsection{Analysis of the Terminal Function $J_T(\mathbf{x})$}

There are several forms of the uncertainty set $\mathcal{P}$. Interested reader can refer to \cite{bgn2009} for a detailed introduction. %We assume that
%$$\mathcal{P}=\{\mathbf{p}\in\mathcal{PV}: \mathbf{p}_0\in\mathcal{PV}, \mathbf{p}-\mathbf{p}_0=\sum_{\ell=1}^L\zeta_{\ell}(y_1, y_2,...,y_m)', \zeta\in \mathfrak{Z}\},$$ where $\mathcal{PV}=\{\mathbf{x}=(x_1,x_2,...,x_m)'\in \mathbb{R}^m: x_i\geq 0, i=1,2,...,m, \sum_{i=1}^mx_i=1\}$ is the probability vector space, $\mathbf{p}_0$ is the nominal operational profile.
Here we provide two types of uncertainty sets, which are commonly used and tractable.

 The first uncertainty set is interval uncertainty set, which can be formulated as
 %$\mathfrak{Z}=\{\zeta\in\mathbb{R}^L: \|\zeta\|_{\infty}=\max_{i=1}^m |\zeta_i| \leq 1\}$. A typical interval uncertainty set is
 $$\mathcal{P}=\{\mathbf{p}\in\mathcal{PV}: \mathbf{l}\leq\mathbf{p}\leq\mathbf{u}\},$$
 where $\mathcal{PV}=\{\mathbf{x}=(x_1,x_2,\cdots,x_m)'\in \mathbb{R}^m: x_i\geq 0, i=1,2,\cdots,m, \sum_{i=1}^mx_i=1\}$ is the probability vector space, and $\mathbf{l}$, $\mathbf{u}\in\mathbb{R}^m$ with $0\leq l_i\leq u_i\leq 1$ for $i=1, 2, \cdots, m$. 
 
 The second uncertainty set is ellipsoidal uncertainty set, which can be formulated as
 %$\mathfrak{Z}=\{\zeta\in\mathbb{R}^L: \|\zeta\|_{2}=\sqrt{\sum_{i=1}^m \zeta_i^2}\leq 1\}$. A typical ellipsoidal uncertainty set is
$$\mathcal{P}=\{\mathbf{p}\in\mathcal{PV}:  (\mathbf{p}-\mathbf{p}_0)' \Sigma^{-1}(\mathbf{p}-\mathbf{p}_0)\leq \varepsilon^2\},$$
where $\Sigma$ is a positive-definite matrix of order $m$,  $\mathbf{p}_0\in\mathcal{PV}$ is the nominal operational profile, and $\varepsilon$ denotes the confidential level of the nominal operational profile. Small $\varepsilon$ implies that the tester is quite sure about the true value of the operational profile.

From (\ref{dp_boundary}) and the fact that $U(x)$ is strictly increasing in $x$, we know that
$$
\arg\min_{\mathbf{p}\in \mathcal{P}} U(R(\mathbf{x},\mathbf{p})) = \arg\min_{\mathbf{p}\in \mathcal{P}} R(\mathbf{x},\mathbf{p})=\arg\min_{\mathbf{p}\in \mathcal{P}}\sum_{i=1}^m p_i(1-\theta_i)^{x_i}.
$$

Therefore, $J_T(\mathbf{x})$ can be obtained by solving
\begin{eqnarray}
\min_{\mathbf{p}}&&\sum_{i=1}^m p_i(1-\theta_i)^{x_i} \label{eq_ter}
\\ \text{subject to}&&   \mathbf{p}\in \mathcal{P}. \nonumber
\end{eqnarray}

Note that the objective function of the above problem is a linear function of $\mathbf{p}$. Thus, if $\mathcal{P}$ is an interval uncertainty set, then the above problem is a linear program; if $\mathcal{P}$ is an ellipsoidal uncertainty set, then it is a quadratically constrained linear program. Each problem can be solved quite efficiently by using commercial solvers such as CPLEX or MOSEK.

In the case of $m=2$, (\ref{eq_ter}) can be solved analytically. In this special case the uncertainty set $\mathcal{P}$ can be written as
$$\mathcal{P}=\{\mathbf{p}=(p_1,p_2)': l_1\leq p_1\leq u_1, p_1+p_2=1\}$$
for some $0\leq l_1\leq u_1\leq 1$. Hence,
\begin{eqnarray*}
&&J_T(\mathbf{x})
\\&=&\min_{l_1\leq p_1\leq u_1}U\left(p_1(1-\theta_1)^{x_1}+p_2(1-\theta_2)^{x_2}\right)
\\ &=& \left\{\begin{array}{ll}U\left(l_1(1-\theta_1)^{x_1}+(1-l_1)(1-\theta_2)^{x_2}\right), &\mbox{if } (1-\theta_1)^{x_1}\geq (1-\theta_2)^{x_2},
\\U\left(u_1(1-\theta_1)^{x_1}+(1-u_1)(1-\theta_2)^{x_2}\right), &\mbox{if }(1-\theta_1)^{x_1}< (1-\theta_2)^{x_2}.\end{array}\right.
\\ &=& \min_{p_1 =l_1, u_1}U\left(p_1(1-\theta_1)^{x_1}+(1-p_1)\cdot(1-\theta_2)^{x_2}\right).
\end{eqnarray*}

Hence, the uncertainty set $\mathcal{P}=\{\mathbf{p}=(p_1,p_2)': l_1\leq p_1\leq u_1, p_1+p_2=1\}$ is equivalent to a two-point set $\mathcal{P}' = \{(l_1,1-l_1)', (u_1,1-u_1)'\}$ in the case of $m=2$.

\subsection{Properties of the Optimal Utility Function $J_t(\mathbf{x})$}

{\proposition\label{pro_4_2} $J_t(\mathbf{x})$ is decreasing in $\mathbf{x}$.}

\begin{proof}
We prove this proposition by backward induction on $t$.

Obviously, from (\ref{eq_reliability}) and (\ref{dp_boundary}) we know that $J_T(\mathbf{x})$ is decreasing in $\mathbf{x}$.

Suppose it holds for $t=k+1$. Next we show that it holds for $t=k$.

By the induction hypothesis, we know that $\mathbb{E}J_{k+1}\left(x_1,x_2,...,x_{i-1}, \sum_{j=1}^{x_i}W_{i,j}, x_{i+1},...,x_m\right)$ is decreasing in $\mathbf{x}$, $i=1,2,...,m$. Thus, from (\ref{dp_recursion}) and the fact that the maximization of decreasing functions is also a decreasing function,  we have $J_{k}(\mathbf{x})$ is also decreasing in $\mathbf{x}$.

Therefore, $J_t(\mathbf{x})$ is decreasing in $\mathbf{x}$.
\end{proof}

This result is consistent with our intuition: With more defects remaining, the smaller the delivered reliability will be. Thus, it is beneficial to remove defects as many as possible.

{\corollary $J_t(\mathbf{x})$ is decreasing in $t$.}

\begin{proof}
From Proposition~\ref{pro_4_2} we know that $\mathbb{E}J_{t+1}\left(x_1,x_2,...,x_{i-1}, \sum_{j=1}^{x_i}W_{i,j}, x_{i+1},...,x_m\right)\geq J_{t+1}(\mathbf{x})$. Thus, from (\ref{dp_recursion}) we know that $J_t(\mathbf{x})\geq J_{t+1}(\mathbf{x})$, which implies that $J_t(\mathbf{x})$ is decreasing in $t$.
\end{proof}

This result is also rather intuitive: More testing time periods remaining implies with more chance the defects will be detected and removed, which means that the delivered reliability will be larger. Thus, a large testing period number $T$ will result in a larger reliability.

{\lemma\label{lemma_1} If $f_i(x)$ is convex in $x$, $i=1,2,...,m$, then $f(x)=\max_{1\leq i\leq m}f_i(x)$ is also convex in $x$.}

\begin{proof}
For any $x_1<x_2$ and $0<\alpha<1$, there exists a number $i_0$ such that $f(\alpha x_1 + (1-\alpha)x_2)=f_{i_0}(\alpha x_1 + (1-\alpha)x_2)$. Thus, we have
\begin{eqnarray*}
&& f(\alpha x_1 + (1-\alpha)x_2)= f_{i_0}(\alpha x_1 + (1-\alpha)x_2)
\\ && \leq \alpha f_{i_0}(x_1)+(1-\alpha) f_{i_0}(x_2) \leq \alpha f(x_1)+(1-\alpha)f(x_2),
\end{eqnarray*}
which implies that $f(x)$ is convex in $x$.
\end{proof}

Note that $x$ can be a vector in the above lemma. By using this lemma, we have the following result.

{\proposition\label{prop_3} In the case that $U(x)=x$ and $\mathcal{P}=\{\mathbf{p}\}$, $J_t(\mathbf{x})$ is convex in $\mathbf{x}$.}
\begin{proof}
We prove this proposition by induction on $t$.

Obviously, $J_T(\mathbf{x})=R(\mathbf{x},\mathbf{p})= \sum_{i=1}^m p_i(1-\theta_i)^{x_i}$ is convex in $\mathbf{x}$.

Suppose it holds for $t=k+1$. We prove it also holds for $t=k$.

By the induction hypothesis, we know that $\mathbb{E}J_{k+1}\left(x_1,x_2,...,x_{i-1}, \sum_{j=1}^{x_i}W_{i,j}, x_{i+1},...,x_m\right)$ is convex in $\mathbf{x}$, $i=1,2,...,m$. Thus, from (\ref{dp_recursion}) and Lemma~\ref{lemma_1} we know that $J_{k}(\mathbf{x})$ is also convex in $\mathbf{x}$.

Therefore, $J_t(\mathbf{x})$ is convex in $\mathbf{x}$.
\end{proof}

$U(x)=x$ means that the tester is risk-neutral and $\mathcal{P}=\{\mathbf{p}\}$ means that the tester is quite sure about the exact value of the user's operational profile. In this case, $J_t(\mathbf{x})$ is convex in $\mathbf{x}$, which implies that $J_t(\mathbf{x}+\mathbf{e}_i)-J_t(\mathbf{x})\geq J_t(\mathbf{x})-J_t(\mathbf{x}-\mathbf{e}_i)$, where $\mathbf{e}_i$ is an $m$-dimensional column vector with all entries 0 except the $i$-th entry 1. Thus, the expected reliability will increase in margin with respect to $x_i$. Note that $J_t(\mathbf{x}+\mathbf{e}_i)-J_t(\mathbf{x})\leq 0$. Thus, if a defect in module $i$ is removed (the defect number in module $i$ changes from $x_i$ to $x_i-1$), the expected reliability can be greatly improved when the defect number in module $i$ (i.e., $x_i$) is small. It seems that it is beneficial to test modules with less defects as removing one defect can significantly improve the expected reliability. However, with less defects remaining in module $i$, one or more defects will be detected and removed with less probability. Thus, it might not be optimal to select the module with minimum number of remaining defects. Similarly, we can argue that it might not be optimal to select the module with maximal number of remaining defects. We will discuss this issue in detail in the following subsection.

Proposition~\ref{prop_3} fails to hold generally if the tester is risk averse or he only has a rough estimate of the users' operational profile $\mathbf{p}$. Next we give two simple examples to illustrate this point.
{\example $m=1$, $U(x)=x-1/2x^2$.}

Obviously, $p=1$ and $U(x)$ is increasing and concave for $0\leq x\leq 1$. We have that $J_T(x)=(1-\theta_1)^x-(1-\theta_1)^{2x}$. Thus, by letting $(1-\theta_1)=e^{-\alpha}$, we have $\partial^2 J_T(x)/\partial x^2=\alpha^2 e^{-\alpha x}(1-2e^{-\alpha x})$, which can not be guaranteed to be nonnegative for all $x\geq 0$ and $0<\alpha$. For instance, when $\alpha=\ln (1.5)$ (i.e., $\theta_1=1/3$), $\partial^2 J_T(x)/\partial x^2<0$ at $x=1$. Thus, $J_T(x)$ is not convex in $x$.

{\example $m=2$, $U(x)=x$, $\mathcal{P}=\{(0.2,0.8)', (0.8, 0.2)'\}$, $\theta_1= 0.3$, $\theta_2 = 0.2$.}

We have that $J_T(12, 19)=0.014$, $J_T(11,19)=0.0155$ and $J_T(13,19)=0.0106$. Note that $J_T(12, 19)>1/2(J_T(11,19)+J_T(13,19))$, which illustrates that $J_T(\mathbf{x})$ is not convex in $\mathbf{x}$.

From the above two examples, we can see that the properties of the optimal utility function depend heavily on the structure of the uncertainty set and the utility function.

\subsection{Analysis of the Optimal Selection Policy}

It is quite intuitive that if it is optimal to test module $i$, it is also optimal when there are more defects in module $i$ given that the numbers of defects remaining in all other modules are unchanged, i.e., the optimal selection policy should have a monotonic property. However, the following discussion shows that the monotonic property of the optimal selection policy fails to hold.

We consider a special case in which $U(x)=x$ and $\mathcal{P}=\{\mathbf{p}\}$. We have $$J_T(\mathbf{x}) = \sum_{i=1}^m p_i (1-\theta_i)^{x_i}.$$

Therefore, it holds that
\begin{eqnarray*}
&& \mathbb{E}J_{T}\left(x_1,x_2,\cdots,x_{i-1}, \sum_{j=1}^{x_i}W_{i,j}, x_{i+1},\cdots,x_m\right)
\\ & = & \sum_{k=0}^{x_i}\left(\begin{array}{c}x_i\\k\end{array}\right) (1-\theta_i)^k \theta_i^{x_i-k} J_{T}(x_1,x_2,\cdots,x_{i-1}, k, x_{i+1},\cdots,x_m)
\\ & = & \sum_{k=0}^{x_i}\left(\begin{array}{c}x_i\\k\end{array}\right) (1-\theta_i)^k \theta_i^{x_i-k} \left(\sum_{j=1, j\neq i}^m p_j (1-\theta_j)^{x_j} + p_i(1-\theta_i)^{k}\right)
\\ & = & \sum_{j=1, j\neq i}^m p_j (1-\theta_j)^{x_j} + p_i\theta_i^{x_i}\sum_{k=0}^{x_i}\left(\begin{array}{c}x_i\\k\end{array}\right) (1-\theta_i)^{2k}\theta_i^{-k}
\\ & = & \sum_{j=1, j\neq i}^m p_j (1-\theta_j)^{x_j} + p_i\theta_i^{x_i}(1 + (1-\theta_i)^2/{\theta_i})^{x_i}
\\ & = & \sum_{j=1, j\neq i}^m p_j (1-\theta_j)^{x_j} + p_i(1-\theta_i + \theta_i^2)^{x_i}.
\end{eqnarray*}

It follows from (\ref{dp_recursion}) that
\begin{eqnarray*}
J_{T-1}(\mathbf{x})&=&\max_{1\leq i\leq m} \left\{\sum_{j=1, j\neq i}^m p_j (1-\theta_j)^{x_j} + p_i(1-\theta_i + \theta_i^2)^{x_i}\right\}
\\ & = & \max_{1\leq i\leq m} p_i((1-\theta_i + \theta_i^2)^{x_i}-(1-\theta_i)^{x_i}) + \sum_{j=1}^m p_j (1-\theta_j)^{x_j}.
\end{eqnarray*}

Hence, at state $(\mathbf{x},T-1)$, it is optimal to select from module $i^*$ with
$$i^* = \arg \max_{1\leq i\leq m} p_i((1-\theta_i + \theta_i^2)^{x_i}-(1-\theta_i)^{x_i}).$$
Note that $p_i((1-\theta_i + \theta_i^2)^{x_i}-(1-\theta_i)^{x_i})$ is not increasing in $x_i$. Thus, it is possible that $p_i((1-\theta_i + \theta_i^2)^{x_i}-(1-\theta_i)^{x_i})>p_j((1-\theta_j + \theta_j^2)^{x_j}-(1-\theta_j)^{x_j})$ for all $j\neq i$ while $p_i((1-\theta_i + \theta_i^2)^{x_i+1}-(1-\theta_i)^{x_i+1})<p_j((1-\theta_j + \theta_j^2)^{x_j}-(1-\theta_j)^{x_j})$ for some $j\neq i$. Therefore, if it is optimal to test module $i$, it might not be optimal to do so when there are more defects in module $i$. The following numerical example justifies our argument.

{\example $m=2$, $U(x)=x$, $\mathcal{P}=\{(0.2,0.8)'\}$, $\theta_1= 0.2$, $\theta_2 = 0.1$, $T=2$.}

In the above example, we find that at the beginning of the testing period, it is optimal to test module $1$ when there are 24 defects remaining in module $1$ and 15 defects remaining in module $2$ while it is optimal to test module $2$ when there are 25 defects remaining in module $1$ and 15 defects remaining in module $2$. This example tells us that the monotonic property of the optimal selection policy fails to hold generally.

\section{Numerical Study}

In this section, we conduct several numerical examples to examine the impact of testing objective, the benefit of robust optimization, and the effects of tester's risk aversion behavior.

\subsection{Impact of Testing Objective}
A lot of literature such as~\cite{c01,c02,c03} discusses how to dynamically select the tested modules to minimize the number of residual defects while this paper's objective is to maximize delivered reliability. For most cases these two objectives are not equivalent. Suppose that the operational profile of the users is $\mathbf{p}=(p_1,p_2,\cdots,p_m)'$ and the number of defects in module $i$ at the end of the testing process is $x_i$, $i=1,2,\cdots,m$. The delivered reliability is $\sum_{i=1}^m p_i(1-\theta_i)^{x_i}$  while the number of residual defects is $\sum_{i=1}^mx_i$. Therefore, these two objectives will be not equivalent.

In the previous literature, the former objective is frequently used~\cite{c01,c02}, i.e., the testing objective is to minimize the number of residual defects. However, the goal of software testing is to improve the delivered reliability, rather than to minimize the residual defect number. Although the residual defect number can be used as an index to measure the reliability, there may be a substantial gap between minimizing the defect number and maximizing the delivered reliability. Some modules may have many defects, but the users seldom use these modules and each defect in these modules can trigger a failure with a relatively small probability. In this situation, the defects in these modules will not count much for the delivered reliability. Therefore, maximizing the delivered reliability is a more preferred objective from the users' perspective. The following is a simple numerical example to illustrate the difference of these two objectives.

{\example $m=2$, $N_1=40$, $N_2=50$, $T=40$, $\theta_1=0.015$, $\theta_2=0.02$, $\mathcal{P}=\{(0.2,0.8)'\}$.}

 If the objective is to maximize expected delivered reliability (assuming that the utility function is $U(x)=x$), the resulting expected reliability is 0.5382. However, if we use minimizing expected defect number as its objective, the resulting expected delivered reliability will be 0.4722, which is significantly less than 0.5382. Thus, these two objectives can result in quite different values of reliability.

\subsection{Necessity of Considering Operational Profile Uncertainty}

Usually different softwares have different operational profiles. The operational profile of a software with similar functionality and the market research may be helpful in obtaining a rough estimate of the operational profile. However, it is difficult, if not impossible, to know the exact value of the operational profile until the software is released to the market. It can be easily foreseen that different operational profile might lead to different optimal selection policy. A selection policy which is optimal for one operational profile may perform quite badly for another operational profile. In the case of unknown operational profile, a selection policy that performs robustly in the operational profile might be preferable.

Next we give a numerical example to show that it is necessary to take uncertainty of operational profile into account during the testing process.

{\example $m=2$, $N_1=40$, $N_2=25$, $T=40$, $\theta_1=0.025$, $\theta_2=0.04$.}

 We assume that the tester's objective is to maximize expected delivered reliability, i.e., he is risk neutral and thus $U(x)=x$. Suppose that the tester cannot know the exact value of $p_1$ (which is 0.5), but he knows that $0.48\leq p_1\leq 0.62$. If the tester knows the exact value of $p_1$, he can use (\ref{dp_boundary}) and (\ref{dp_recursion}) to obtain a selection policy with the expected reliability $0.4809$. However, the exact value of $p_1$ is unknown to him. If the tester uses robust optimization technique to find the testing policy, the resulting reliability is $0.477$. The gap which is defined to be the relative error of the two reliabilities is $0.815\%$. If the tester mistakenly choose the value of $p_1$ to be $p_w$, he will obtain a selection policy which is different with the optimal way and generates a lower expected reliability. Let $p_w$ vary from $0.48$ to $0.62$, we obtain the gaps of the expected reliability as shown in Table~\ref{tab_1}.

\begin{table}
\caption{The gaps of the expected reliability as the true value of $p_1$ varies}\label{tab_1}
\begin{tabular}{|c|c|c|c|c|c|c|c|c|}
  \hline
  $p_1$ & 0.48 & 0.50 & 0.52 & 0.54 & 0.56 & 0.58 & 0.60 & 0.62 \\
  Gap & 0.0641\% & 0 & 0.11\% & 0.3642\% & 0.6738\% & 1.302\% & 2.1970\% & 3.3791\%  \\
  \hline
\end{tabular}
\end{table}

From Table~\ref{tab_1} we find that the gaps are quite sensitive to the value of $p_w$. If the tester mistakenly choose the value of $p_1$ to be 0.62, he will obtain a expected reliability which is significantly lower than the optimal expected reliability. However, if he considers the uncertainty of operational profile by using robust optimization method, the resulting reliability will be globally satisfying.

\subsection{Impact of Risk Aversion Behavior}
In this section, we consider a special class of utility functions to examine the impact of the tester's risk aversion behaviors on the reliability. Specifically, the utility function of the tester is assumed to be $U(x)=1-\exp(-x/\gamma)$, where $\gamma$ is the tester's risk tolerance. Larger value of $\gamma$ implies that the tester has more risk tolerance to the delivered reliability. It can be easily seen that the tester tends to be risk neutral as $\gamma$ goes to infinity. In fact, $\gamma=+\infty$ corresponds to the case that the tester is risk neutral.

Next we give a numerical example to show the impact of the tester's risk aversion behaviors on the reliability.

{\example $m=2$, $N_1 = 30$, $N_2 = 20$, $\mathcal{P}=\{(0.4,0.6)'\}$, $\theta_1 = 0.1$, $\theta_2 = 0.2$ and $T=15$.}

\begin{figure}
  \includegraphics[width=12cm]{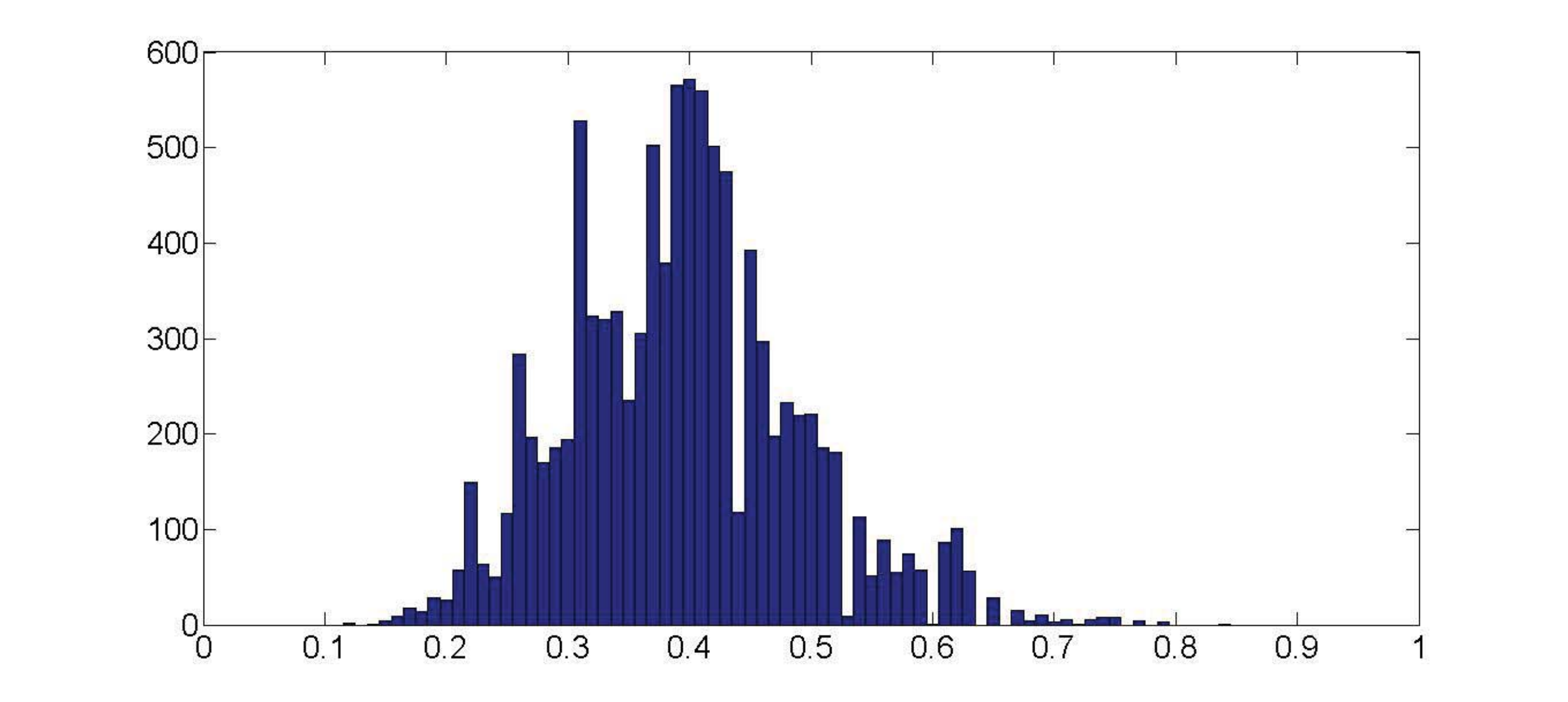}\\
  \caption{$\gamma=0.001$}\label{fig_1}
\end{figure}

\begin{figure}
  \includegraphics[width=12cm]{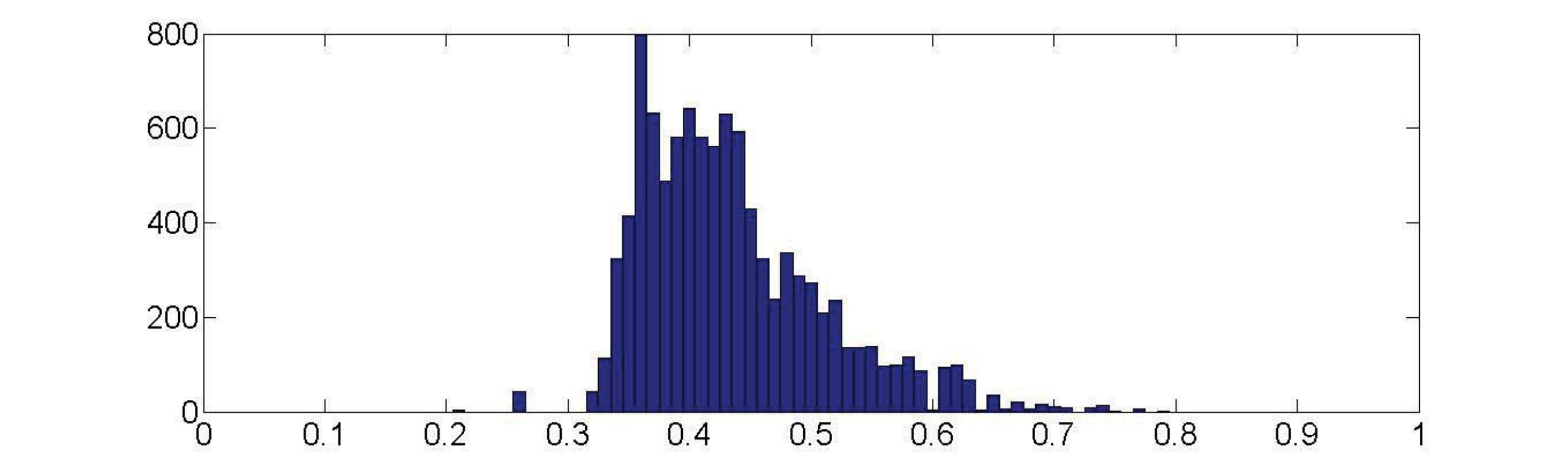}\\
  \caption{$\gamma=0.01$}\label{fig_2}
\end{figure}

\begin{figure}
  \includegraphics[width=12cm]{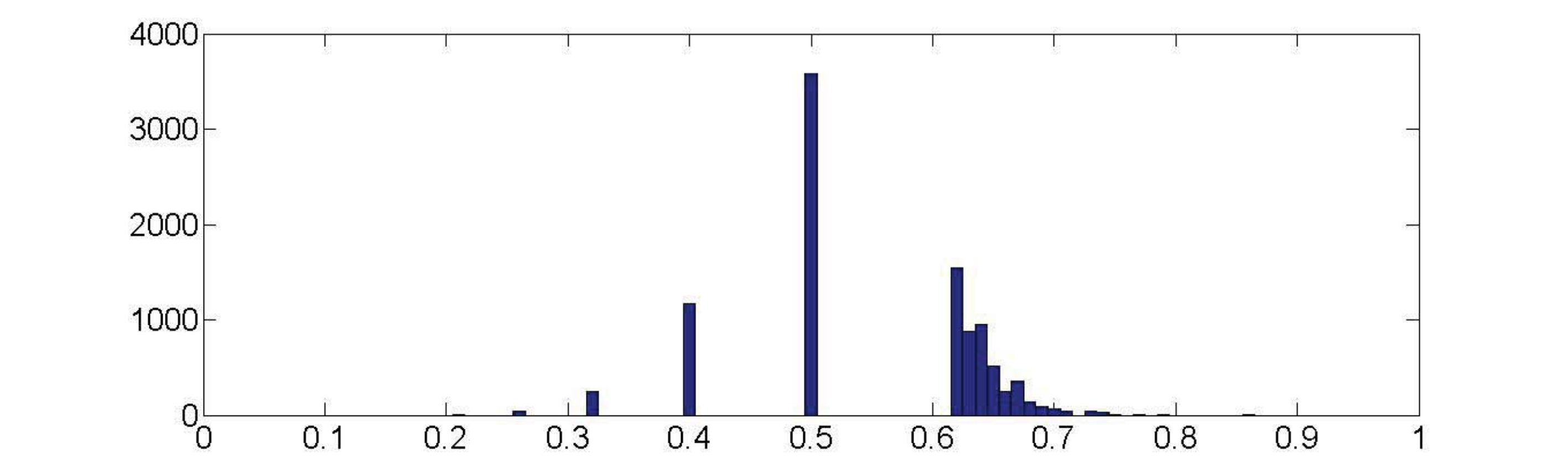}\\
  \caption{$\gamma=0.1$}\label{fig_3}
\end{figure}

\begin{figure}
  \includegraphics[width=12cm]{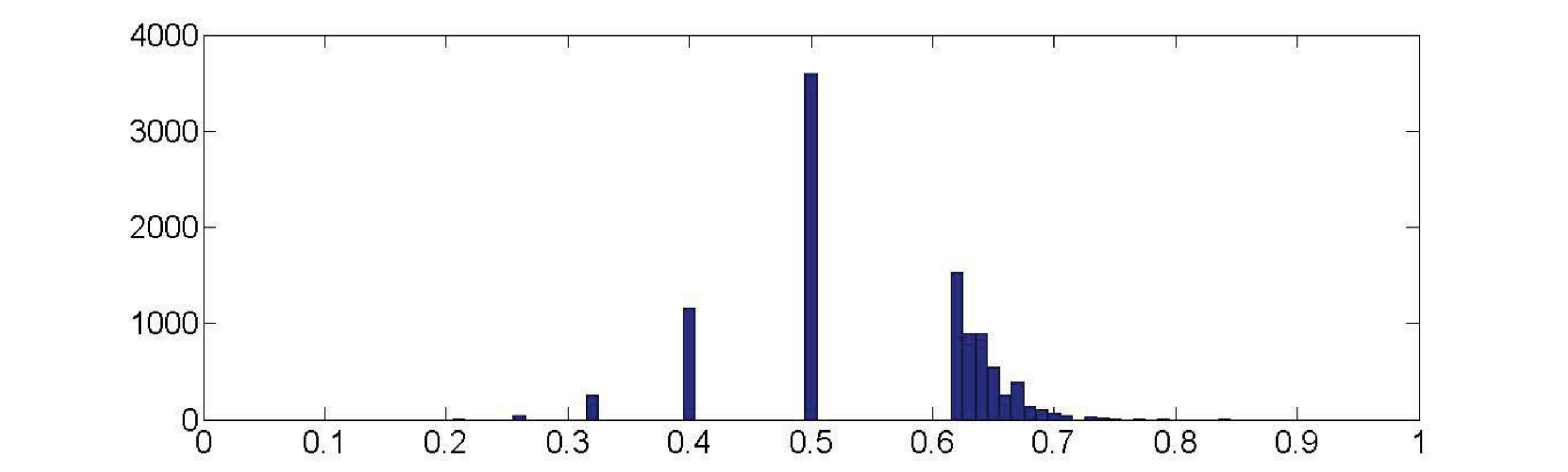}\\
  \caption{$\gamma=1$}\label{fig_4}
\end{figure}

We obtain the optimal selection policies for $\gamma=0.001, 0.01, 0.1$ and $1$ by solving (\ref{dp_boundary}) and (\ref{dp_recursion}), respectively. Then, we run 10000 simulations according to these obtained selection policies to obtain the distribution of the delivered reliability. For instance, Figures~\ref{fig_1} and \ref{fig_2} are histograms recording the frequency of the delivered reliability for $\gamma=0.001$ and $0.01$, respectively. They show a relatively uniform distribution of the delivered reliability. In contrast, Figures~\ref{fig_3} and~\ref{fig_4} show a relatively scattered distribution of the delivered reliability.

Table~\ref{tab_2} shows the average reliability and the variance for $\gamma=0.001, 0.01, 0.1$ and $1$, respectively. It shows that small value of $\gamma$ will result in a small average reliability but the resulting variance is not always small. This is justified by Figures \ref{fig_1}-\ref{fig_4}: Figure~\ref{fig_1} and Figure~\ref{fig_2} show that the reliability is quite centred around the average reliability. Figure~\ref{fig_3} and Figure~\ref{fig_4} show that the reliability is quite deviated from the average reliability (noting that the average reliability is around 0.55).

\begin{table}
\caption{The average reliability and the variance}\label{tab_2}
\begin{center}
\begin{tabular}{|c| c| c| c| c|}
\hline
$\gamma$ & 0.001 & 0.01 & 0.1 & 1
\\
\hline
Average & 0.3947 & 0.4346 & 0.5504 & 0.5512
\\
Variance & 0.0087 & 0.0057 & 0.0099 & 0.0099
\\
\hline
\end{tabular}
\end{center}
\end{table}

%\subsection{Comparison of Optimal Selection Policy and Heuristic Selection Policies}
%
%In practical testing process, the tester may use heuristic selection policies rather than the optimal selection policy. The main reason is that the optimal selection policy is hard to obtain and it depends on the model parameters $N_i$, $\theta_i$ which is unknown in reality and needs to be estimated. However, the tester may not want to put much effort into estimating parameters and obtaining optimal policy. In this section, we propose two heuristic policies and compare their performance with the optimal policy.
%
%
%
%Let $A_t$ be the $t$-th testing action taken since the beginning of software testing, and $A_t=i$ means that the $t$-th testing action picks up a test case from module $C_i$. If for each testing, $\mathbb{P}(A_t=i)=q_i$, where $\sum_{i=1}^mq_i=1$, $q_i\geq 0$, $i=1,2,...,m$, $t=1,2,...,T$, this corresponds to \textit{random testing}~\cite{ai12}. If the sequence $\{A_1,A_2,..., A_T\}$ forms a Markov chain with $\mathbb{P}(A_{t+1}=l|A_t=k)=q_{kl}$, $t=0,1,...,T-1$, where $q_{kl}$ is the transition probability from action $k$ to action $l$, satisfying $q_{kl}\geq 0$, $\sum_{l=1}^mq_{kl}=1$, $k,l=1,2,\cdots,m$, then this corresponds to \textit{Markov usage model based testing}\cite{wt94}.

\section{Conclusion}

This paper presents a new framework to test case selection problem of how to select test cases to maximize delivered software reliability.  Typically, the goal of testing is eliminating as many faults as possible, under the constraints of cost and time.  One main reason of not using software reliability as the testing goal is that it cannot be  estimated accurately during the testing process.  More precisely, software reliability is defined as the probability of encountering a failure (not a fault) in specified operational profile, while the operational profile is difficult to estimate prior to release.  We address this problem and treat the uncertainty of operational profile by using robust optimization. Moreover, considering that the realized reliability can be quite different from the expected reliability, the uncertainty (variability) of the reliability is also taken into account.  This kind of variability implies risk for the tester. Hence, we use a strictly increasing and concave utility function to reflect the tester's risk aversion behavior. In order to study the impact of testing objective, the benefit of robust optimization, and the effects of the tester's risk aversion behavior, we conduct several numerical examples, finding that testing objective has a significant effect on the delivered reliability.

There are at least fourth topics worthy of future research. First, in this paper the length of testing time periods $T$ is fixed. However, during the actual testing activities, the tester might decide to release the software earlier or later based on the testing status quo. That is, the tester can also dynamically adjust the testing time length according to the testing process. Second, in this paper we present the testing time length $T$ as the time limitation while there are other limitations such as testing expenses which have not been considered yet. Third, it is assumed that the defect number in each module at the beginning of the testing process $N_i$ and the detect probability $\theta_i$ are known, which does not meet the reality. These parameters must be estimated during the testing process. In order to incorporate parameter estimation into the testing process, adaptive control might be used to model this problem and Bayesian dynamic program might be a useful analytical tool. Finally, when the number of tested modules $m$ is large, it will take a lot of time to compute (\ref{dp_recursion}) directly. This problem is especially acute for a complex software system with hundreds of modules. We might need approximate dynamic programming method to tackle this problem. In this case, heuristic selection policies might be more preferable.

\section*{Acknowledgements}

The authors gratefully acknowledge the partial support of the National Natural Science Foundation
of China (NSFC) under grants 71401159, 71390334, 11271356, 61272164 and the Fundamental
Research Funds for the Central Universities under grant WK2040160009.

\end{document}